\documentclass[review]{elsarticle}
\usepackage{graphicx}% Include figure files
\usepackage{dcolumn}% Align table columns on decimal point
\usepackage{bm}% bold math
\usepackage{longtable}
\usepackage{amsmath}
\usepackage{amsfonts,amssymb}
\usepackage{mathtools,amsthm,adjustbox}
\usepackage{lipsum}
\usepackage[margin=2.9cm]{geometry}
\def\bn{\begin{equation}}
\def\en{\end{equation}}
\def\bny{\begin{eqnarray}}
\def\eny{\end{eqnarray}}
\def\be{\begin{eqnarray*}}
\def\ee{\end{eqnarray*}}
\def\bc{\begin{center}}
\def\ec{\end{center}}

\def\({\left(}
\def\){\right  )}
\def\[{\left[}
\def\]{\right]}
\def\bc{\begin{center}}
\def\ec{\end{center}}

\newtheorem{theorem}{Theorem}

\def\bn{\begin{equation}}
\def\en{\end{equation}}
\def\bny{\begin{eqnarray}}
\def\eny{\end{eqnarray}}
\def\be{\begin{eqnarray*}}
\def\ee{\end{eqnarray*}}
\def\bdn{\begin{dfn}}
\def\edn{\end{dfn}}
\def\btm{\begin{thm}}
\def\etm{\end{thm}}
\def\bpf{\begin{proof}}
\def\epf{\end{proof}}
\def\bpn{\begin{pro}}
\def\epn{\end{pro}}
\def\brk{\begin{rem}}
\def\erk{\end{rem}}
\def\bcy{\begin{cor}}
\def\ecy{\end{cor}}
\def\blm{\begin{lem}}
\def\elm{\end{lem}}
\def\bex{\begin{exm}}
\def\eex{\end{exm}}

\def\dsum{\displaystyle\sum}

\usepackage{lineno,hyperref}
\modulolinenumbers[5]

\usepackage{numcompress}

\begin{document}

\begin{frontmatter}

\title{Approximate Noether Symmetries and Collineations for Regular
Perturbative Lagrangians}

\author{Andronikos Paliathanasis}
 \address{Instituto de Ciencias F\'{\i}sicas y Matem\'{a}ticas, Universidad Austral de
Chile, Valdivia, Chile\\ Institute of Systems Science, Durban University of Technology, 
Durban, South Africa}
\ead{ anpaliat@phys.uoa.gr}

\author{Sameerah Jamal \corref{mycorrespondingauthor}}
\address{School of Mathematics and Centre for Differential Equations,Continuum Mechanics and Applications,
University of the Witwatersrand, Johannesburg, South Africa}
\ead{Sameerah.Jamal@wits.ac.za}
\cortext[mycorrespondingauthor]{Corresponding author}

\begin{abstract}
Regular perturbative Lagrangians that admit approximate Noether symmetries and
approximate conservation laws are studied. Specifically, we investigate the
connection between approximate Noether symmetries and collineations of the
underlying manifold. In particular we determine the generic Noether symmetry
conditions for the approximate point symmetries and we find that for a class
of perturbed Lagrangians, Noether symmetries are related to the elements of
the Homothetic algebra of the metric which is defined by the unperturbed
Lagrangian. Moreover, we discuss how exact symmetries become approximate
symmetries. Finally, some applications are presented.\end{abstract}


\end{frontmatter}

%\linenumbers

\section{Introduction}

Symmetries play an important role in the study of differential equations.
The existence of a symmetry vector implies the existence of a transformation
which reduces the order of the differential equation (for ordinary
differential equations) or the number of dependent variables (for partial
differential equations). However, the existence of a symmetry vector 
indicates that there exists a curve in the phase space of the dynamical system
which constrains the solution of the differential equation. This specific
curve is a conservation law for the differential equation.

There are various ways to construct conservation laws, for instance see \cite%
{con1,con2,con3,con4}. One of the most well-known, and simplest methods
for the determination of conservation laws is the application of Noether's
theorems \cite{noet}. In particular, the first Noether's theorem states that, if
the Lagrangian function which describes a dynamical system changes under the
action of a point transformation such that the Action integral is
invariant, the dynamical system is also invariant under the action of the
same point transformation. Moreover, a conservation law corresponds to this
point transformation according to Noether's second theorem.

Usually, when we refer to symmetries, we consider the exact symmetries.
However, for perturbative dynamical systems the context of symmetries is
extended and the so-called approximate symmetries are defined \cite{bai,
fush, pek,gaz1,gaz2,gaz3,gaz4,leach1}. In this work we are interested in the
application of Noether's theorem for approximate symmetries on some regular
perturbative Lagrangians. Approximate Noether symmetries \cite{gov,ham1}
provide approximate first integrals, functions which can be used as
conservation laws until a specific step in numerical integrations. This
kind of approximate conservation laws have  played an important role for
the study of chaotic systems - for an extended  discussion we refer the
reader to an application in Galactic dynamics \cite{gal1,gal2,gal3}.

Whilst recently, the advent of automated software algorithms has made light
work of calculating symmetries \cite{nucc11}. Such programs are often
limited by models involving many variables or higher-order perturbations.
This problem, in part, has fueled the need to write this paper. Here, we
take a compound problem, whereby scientists have previously relied on numerical
techniques for analysis, and instead frame it in the context of an analytical scheme. We
present a set of conditions that may be specialized for appropriate
Lagrangian functions that necessarily contains a perturbation. 
Inspired by the approach of Tsamparlis and Paliathanasis \cite%
{tsam1,tsam2,tsam3},  we show how those conditions can be solved with the use
of some theorems from differential geometry. 
Indeed, geometric based theories have far reaching applications \cite{sit, boz, j10}.

Specifically, in this paper, the Noether conditions, or symmetry determining system of
equations, are formulated by contemplating point transformations in
ascending order of the perturbation parameter $\varepsilon$. To illustrate
the advantages of such a formulation, the general conditions are applied to
the perturbations of oscillator type equations corresponding to $n$
dimensions. Moreover, we discuss the admitted approximate conserved
quantities for symmetries of first-order up-to $n^{th}$-order.

This paper assumes familiarity with symmetry-based methods and so does not
add to the volume of the work by recapitulating the well established theory.
However, in stating this we recommend the interested reader to consult the
books \cite{o} and \cite{hs}, which contain various aspects concerning Lie
and Noether symmetries.

The objective of this paper is two-fold. First, we  utilize a
generalized Lagrangian to formulate Noether symmetry conditions, for
symmetries of higher-order perturbations. Thereafter, the latter is
used to establish the corresponding approximate conservation laws, also of
higher-order perturbations, via Noether's theorem. In this regard, to cope
with the complexity of our derivation, we employ the Einstein summation
convention, and require that indices enclosed in parentheses indicate
symmetrization, for instance, $F_{(ij)}=\frac{1}{2}(F_{ij}+F_{ji}).$

The family of perturbed Lagrangians that we assume are%
\begin{equation}
L\left( t,x^{k},\dot{x}^{k},\varepsilon \right) =L_{0}\left( t,x^{k},\dot{x}%
^{k}\right) +\varepsilon L_{1}\left( t,x^{k},\dot{x}^{k}\right) +O\left(
\varepsilon ^{2}\right) ,  \label{ANS.01a}
\end{equation}%
where we stipulate that the exact and approximate terms are defined by the
regular Lagrangians 
\begin{eqnarray}
L_{0}\left( t,x^{k},\dot{x}^{k}\right)  &=&\frac{1}{2}g_{ij}\dot{x}^{i}\dot{x%
}^{j}-V_{0}\left( t,x^{k}\right) ,  \label{ANS.05} \\
L_{1}\left( t,x^{k},\dot{x}^{k}\right)  &=&\frac{1}{2}h_{ij}\dot{x}^{i}\dot{x%
}^{j}-V_{1}\left( t,x^{k}\right) ,  \label{ANS.06}
\end{eqnarray}%
respectively, where $g_{ij}=g_{ij}\left( x^{k}\right) ,h_{ij}=h_{ij}\left(
x^{k}\right) $~and dot denotes total derivative with respect to the independent
parameter $t$, i.e. $\dot{x}^{i}=\frac{dx^{i}}{dt}$. Such Lagrangians are
not limited by any specific assumptions in order to preserve generality. The
plan of the paper is as follows.

Section \ref{section2} is the main core of our analysis where we derive the
approximate Noether symmetry conditions for regular Lagrangians of order $%
\left( \varepsilon ^{1}\right) $ and $\left( \varepsilon ^{n}\right) $ in
general. Moreover, from the conditions we see that there exists a link
between the approximate symmetries and the Homothetic vector fields of the
metric tensor $g_{ij}$. In Section \ref{section4} we demonstrate our results
by applying our approach to various examples of maximally symmetric systems,
and also on a $sl\left( 2,R\right) $ exact invariant system. For each case we
derive the approximate Noether symmetries and the corresponding approximate
conservation laws. Finally, in Section \ref{section5} we discuss our results
and draw our conclusions.

\section{Approximate Noether Conditions}

\label{section2}

In the interest of clarity and completeness we have decided to present the
immediate work that follows in great detail, dividing the procedure into
important steps. Following Govinder et al. \cite{gov}, under a point
transformation%
\begin{eqnarray}
\bar{x}^{i} &=&x^{i}+a\left( \eta _{\left( 0\right) }^{i}+\varepsilon \eta
_{\left( 1\right) }^{i}+O\left( \varepsilon ^{2}\right) \right) ,
\label{ANS.02} \\
\bar{t} &=&t+a\left( \xi _{\left( 0\right) }+\varepsilon \xi _{\left(
1\right) }+O\left( \varepsilon ^{2}\right) \right) ,  \label{ANS.03}
\end{eqnarray}%
with $\xi _{A}=\xi _{A}\left( t,x^{k}\right) ~,~\eta _{A}^{i}=\eta
_{A}^{i}\left( t,x^{k}\right) ,~A=0,1$ and $``a,\varepsilon "$ as two
infinitesimal parameters, we have the generator%
\begin{equation}
X=X_{0}+\varepsilon X_{1}+O\left( \varepsilon ^{2}\right) ,  \label{ANS.G}
\end{equation}%
where~$X_{A}=\xi _{A}\partial _{t}+\eta _{A}^{i}\partial _{i}$. Here, $X$ is
a first-order approximate vector field composed of an exact ($X_{0}$) and an
approximate ($X_{1}$) part. The generator (\ref{ANS.G}) is a Noether point
symmetry if the following equation is satisfied%
\begin{equation}
X^{[1]}L+L\frac{d\xi }{dt}=\frac{df}{dt},
\end{equation}%
or equivalently 
\begin{equation}
\left( X_{0}^{\left[ 1\right] }+\varepsilon X_{1}^{\left[ 1\right] }\right)
\left( L_{0}+\varepsilon L_{1}\right) +\left( L_{0}+\varepsilon L_{1}\right) 
\frac{d}{dt}\left( \xi _{0}+\varepsilon \xi _{1}\right) -\frac{d}{dt}\left(
f_{0}+\varepsilon f_{1}\right) =O\left( \varepsilon ^{2}\right) ,
\label{ANS.04}
\end{equation}%
where $f_{A}=f_{A}\left( t,x^{i}\right) $ and the term $X_{A}^{\left[ 1%
\right] }$ is the first prolongation expressed as 
\begin{equation}
X_{A}^{\left[ 1\right] }=\xi _{A}\partial _{t}+\eta _{A}^{i}\partial
_{i}+\left( \dot{\eta}_{A}^{i}-\dot{x}^{i}\dot{\xi}_{A}\right) \partial _{%
\dot{x}^{i}}.
\end{equation}

Inserting (\ref{ANS.05}) and (\ref{ANS.06}) into the left-hand-side of the
Noether condition (\ref{ANS.04}), we find 
\begin{eqnarray*}
X_{0}^{\left[ 1\right] }L_{0} &=&\left( \xi _{0}\partial _{t}+\eta
_{0}^{k}\partial _{k}+\left( \dot{\eta}_{0}^{k}-\dot{x}^{k}\dot{\xi}%
_{0}\right) \partial _{\dot{x}^{k}}\right) \left( \frac{1}{2}g_{ij}\dot{x}%
^{i}\dot{x}^{j}-V_{0}\left( t,x^{k}\right) \right) \\
&=&\frac{1}{2}g_{ij,k}\eta _{0}^{k}\dot{x}^{i}\dot{x}^{j}+g_{ij}\left( \dot{%
\eta}_{0}^{i}-\dot{x}^{i}\dot{\xi}_{0}\right) \dot{x}^{j}-\xi
_{0}V_{0\,,t}-\eta _{0}^{k}V_{0,i} \\
&=&\frac{1}{2}g_{ij,k}\eta _{0}^{k}\dot{x}^{i}\dot{x}^{j}+g_{ij}\left( \eta
_{0,k}^{i}-\xi _{0,t}\right) \dot{x}^{j}\dot{x}^{k}+g_{ij}\eta _{0,t}^{i}%
\dot{x}^{j}-g_{ij}\xi _{0,k}\dot{x}^{i}\dot{x}^{j}\dot{x}^{k} \\
&&-\xi _{0}V_{0\,,t}-\eta _{0}^{k}V_{0,i}.
\end{eqnarray*}

Similarly, 
\begin{eqnarray*}
\varepsilon X_{0}^{\left[ 1\right] }L_{1} &=&\varepsilon \Bigg(\frac{1}{2}%
h_{ij,k}\eta _{0}^{k}\dot{x}^{i}\dot{x}^{j}+h_{ij}\left( \eta _{0,k}^{i}-\xi
_{0,t}\right) \dot{x}^{j}\dot{x}^{k}+h_{ij}\eta _{0,t}^{i}\dot{x}^{j} \\
&&\phantom{aaaaaaaaaaaaaaaa}-h_{ij}\xi _{0,k}\dot{x}^{i}\dot{x}^{j}\dot{x}%
^{k}-\xi _{0}V_{1\,,t}-\eta _{0}^{k}V_{1,i}\Bigg), \\
\varepsilon X_{1}^{\left[ 1\right] }L_{0} &=&\varepsilon \Bigg(\frac{1}{2}%
g_{ij,k}\eta _{1}^{k}\dot{x}^{i}\dot{x}^{j}+g_{ij}\left( \eta _{1,k}^{i}-\xi
_{1,t}\right) \dot{x}^{j}\dot{x}^{k}+g_{ij}\eta _{1,t}^{i}\dot{x}^{j} \\
&&\phantom{aaaaaaaaaaaaaaaa}-g_{ij}\xi _{1,k}\dot{x}^{i}\dot{x}^{j}\dot{x}%
^{k}-\xi _{1}V_{0\,,t}-\eta _{1}^{k}V_{0,i}\Bigg), \\
\varepsilon ^{2}X_{1}^{\left[ 1\right] }L_{1} &=&\varepsilon ^{2}\Bigg(\frac{%
1}{2}h_{ij,k}\eta _{1}^{k}\dot{x}^{i}\dot{x}^{j}+h_{ij}\left( \eta
_{1,k}^{i}-\xi _{1,t}\right) \dot{x}^{j}\dot{x}^{k}+h_{ij}\eta _{1,t}^{i}%
\dot{x}^{j} \\
&&\phantom{aaaaaaaaaaaaaaaa}-h_{ij}\xi _{1,k}\dot{x}^{i}\dot{x}^{j}\dot{x}%
^{k}-\xi _{1}V_{1\,,t}-\eta _{1}^{k}V_{1,i}\Bigg).
\end{eqnarray*}

For the middle terms, we have the relation 
\begin{equation*}
\left( L_{0}+\varepsilon L_{1}\right) \frac{d}{dt}\left( \xi
_{0}+\varepsilon \xi _{1}\right) =\dot{\xi}_{0}L_{0}+\varepsilon L_{1}\dot{%
\xi}_{0}+\varepsilon \dot{\xi}_{1}L_{0}+\varepsilon ^{2}\dot{\xi}_{1}L_{1},
\end{equation*}%
and therefore
\begin{eqnarray*}
\dot{\xi}_{0}L_{0} &=&\left( \xi _{0,t}+\xi _{0,k}\dot{x}^{k}\right) \left( 
\frac{1}{2}g_{ij}\dot{x}^{i}\dot{x}^{j}-V_{0}\left( t,x^{k}\right) \right) \\
&=&\frac{1}{2}\xi _{0,t}g_{ij}\dot{x}^{i}\dot{x}^{j}+\frac{1}{2}\xi
_{0,k}g_{ij}\dot{x}^{i}\dot{x}^{j}\dot{x}^{k}-\xi _{0,t}V_{0}-V_{0}\xi _{0,k}%
\dot{x}^{k}.
\end{eqnarray*}

Similarly, the remaining middle terms are 
\begin{eqnarray*}
\varepsilon L_{1}\dot{\xi}_{0} &=&\varepsilon \left( \frac{1}{2}\xi
_{0,t}h_{ij}\dot{x}^{i}\dot{x}^{j}+\frac{1}{2}\xi _{0,k}h_{ij}\dot{x}^{i}%
\dot{x}^{j}\dot{x}^{k}-\xi _{0,t}V_{1}-V_{1}\xi _{0,k}\dot{x}^{k}\right) , \\
\varepsilon \xi _{1}L_{0} &=&\varepsilon \left( \frac{1}{2}\xi _{1,t}g_{ij}%
\dot{x}^{i}\dot{x}^{j}+\frac{1}{2}\xi _{1,k}g_{ij}\dot{x}^{i}\dot{x}^{j}\dot{%
x}^{k}-\xi _{1,t}V_{0}-V_{0}\xi _{1,k}\dot{x}^{k}\right) , \\
\varepsilon ^{2}\xi _{1}L_{1} &=&\varepsilon \left( \frac{1}{2}\xi
_{1,t}h_{ij}\dot{x}^{i}\dot{x}^{j}+\frac{1}{2}\xi _{1,k}h_{ij}\dot{x}^{i}%
\dot{x}^{j}\dot{x}^{k}-\xi _{1,t}V_{1}-V_{1}\xi _{1,k}\dot{x}^{k}\right) .
\end{eqnarray*}

Lastly, the end terms produce the expression 
\begin{equation*}
\dot{f}_{0}+\varepsilon \dot{f}_{1}=\left( f_{0,t}+f_{0,k}\dot{x}^{k}\right)
+\varepsilon \left( f_{1,t}+f_{1,k}\dot{x}^{k}\right) .
\end{equation*}

Next, in collating terms sans $\varepsilon $ we obtain the expression 
\begin{eqnarray*}
\varepsilon ^{0} &:&\frac{1}{2}g_{ij,k}\eta _{0}^{k}\dot{x}^{i}\dot{x}%
^{j}+g_{ij}\left( \eta _{0,k}^{i}-\xi _{0,t}\right) \dot{x}^{j}\dot{x}%
^{k}+g_{ij}\eta _{0,t}^{i}\dot{x}^{j}-g_{ij}\xi _{0,k}\dot{x}^{i}\dot{x}^{j}%
\dot{x}^{k}-\xi _{0}V_{0\,,t} \\
&&-\eta _{0}^{k}V_{0,i}+\left( \frac{1}{2}\xi _{0,t}g_{ij}\dot{x}^{i}\dot{x}%
^{j}+\frac{1}{2}\xi _{0,k}g_{ij}\dot{x}^{i}\dot{x}^{j}\dot{x}^{k}-\xi
_{0,t}V_{0}-V_{0}\xi _{0,k}\dot{x}^{k}\right) \\
&&-\left( f_{0,t}+f_{0,k}\dot{x}^{k}\right) .
\end{eqnarray*}

By collecting terms of equal powers of $(\dot{x})^{K}$, we find the following
determining system of equations 
\begin{eqnarray}
&&\xi _{0,k}g_{ij}=0,  \label{j1} \\
&&\frac{1}{2}g_{ij,k}\eta _{0}^{k}+g_{k(i}\left( \eta _{0,j)}^{k}-\frac{1}{2}%
\xi _{0,t}\right) =0,\quad g_{ij}\eta _{0,t}^{i}-f_{0,j}-V_{0}\xi _{0,j}=0,
\label{j2} \\
&&\xi _{0}V_{0\,,t}+\eta _{0}^{i}V_{0,i}+\xi _{0,t}V_{0}+f_{0,t}=0,
\label{j3}
\end{eqnarray}%
which reveals that $\xi _{0}=\xi _{0}\left( t\right) $.

Terms involving $\varepsilon ^{1}$ are 
\begin{eqnarray*}
\varepsilon ^{1} &:&\frac{1}{2}h_{ij,k}\eta _{0}^{k}\dot{x}^{i}\dot{x}%
^{j}+h_{ij}\left( \eta _{0,k}^{i}-\xi _{0,t}\right) \dot{x}^{j}\dot{x}%
^{k}+h_{ij}\eta _{0,t}^{i}\dot{x}^{j}-\xi _{0}V_{1\,,t}-\eta _{0}^{k}V_{1,i}
\\
&&+\frac{1}{2}g_{ij,k}\eta _{1}^{k}\dot{x}^{i}\dot{x}^{j}+g_{ij}\left( \eta
_{1,k}^{i}-\xi _{1,t}\right) \dot{x}^{j}\dot{x}^{k}+g_{ij}\eta _{1,t}^{i}%
\dot{x}^{j}-g_{ij}\xi _{1,k}\dot{x}^{i}\dot{x}^{j}\dot{x}^{k}-\xi
_{1}V_{0\,,t} \\
&&-\eta _{1}^{k}V_{0,i}+\left( \frac{1}{2}\xi _{0,t}h_{ij}\dot{x}^{i}\dot{x}%
^{j}-\xi _{0,t}V_{1}-V_{1}\xi _{0,k}\dot{x}^{k}\right) \\
&&+\left( \frac{1}{2}\xi _{1,t}g_{ij}\dot{x}^{i}\dot{x}^{j}+\frac{1}{2}\xi
_{1,k}g_{ij}\dot{x}^{i}\dot{x}^{j}\dot{x}^{k}-\xi _{1,t}V_{0}-V_{0}\xi _{1,k}%
\dot{x}^{k}\right) -\left( f_{1,t}+f_{1,k}\dot{x}^{k}\right) .
\end{eqnarray*}

As before, these expressions are substituted back into the Noether condition
to check monomials in $(\dot{x})^K$, and we have the determining system of
equations, 
\begin{equation}
g_{ij}\xi _{1,k}=0,  \label{j4}
\end{equation}%
which immediately implies that $\xi _{1}=\xi _{1}\left( t\right) $, and
additionally 
\begin{eqnarray}
&\left[ \frac{1}{2}h_{ij,k}\eta _{0}^{k}+h_{k(j}\left( \eta _{0,j)}^{k}-%
\frac{1}{2}\xi _{0,t}\right) \right] +\left[ \frac{1}{2}g_{ij,k}\eta
_{1}^{k}+g_{k(j}\left( \eta _{1,i)}^{k}-\frac{1}{2}\xi _{1,t}\right) \right]%
=0,  \label{j5} \\
&h_{ij}\eta _{0,t}^{i}+g_{ij}\eta _{1,t}^{i}-f_{1,j}=0,  \label{j6} \\
&\xi _{0}V_{1\,,t}+\eta _{0}^{i}V_{1,i}+\xi _{1}V_{0\,,t}+\eta
_{1}^{k}V_{0,i}+\xi _{0,t}V_{1}+\xi _{1,t}V_{0}+f_{1,t}=0.  \label{j7}
\end{eqnarray}%
Note that we impose the restriction that terms involving $\varepsilon ^{2}$
are taken to be zero.

Finally, the system of equations (\ref{j1})-(\ref{j3}) and (\ref{j4})-(\ref%
{j7}) provide the approximate Noether symmetry conditions for the perturbed
Lagrangian (\ref{ANS.01a}) together with equations (\ref{ANS.05}), (\ref%
{ANS.06}) and symmetry generator (\ref{ANS.G}), viz.

$\left( \varepsilon \right) ^{0}:$%
\begin{eqnarray}
L_{\eta _{0}}g_{ij} &=&2\left( \frac{1}{2}\xi _{0,t}\right) g_{ij},
\label{ANS.07} \\
g_{ij}\eta _{0,t}^{i} &=&f_{0,j},  \label{ANS.08} \\
L_{\eta _{0}}V_{0}+\xi _{0,t}V_{0}+\xi _{0}V_{0\,,t}+f_{0,t} &=&0,
\label{ANS.09}
\end{eqnarray}

$\left( \varepsilon \right) ^{1}:$%
\begin{eqnarray}
L_{\eta _{0}}h_{ij}+L_{\eta _{1}}g_{ij}-2\left( \frac{1}{2}\xi _{0,t}\right)
h_{ij}-2\left( \frac{1}{2}\xi _{1,t}\right) g_{ij} &=&0,  \label{ANS.10} \\
h_{ij}\eta _{0,t}^{i}+g_{ij}\eta _{1,t}^{i} &=&f_{1,j},  \label{ANS.11} \\
L_{\eta _{0}}V_{1}+L_{\eta _{1}}V_{0}+\xi _{0}V_{1\,,t}+\xi _{0,t}V_{1}+\xi
_{1}V_{0\,,t}+\xi _{1,t}V_{0} &=&-f_{1,t},  \label{ANS.12}
\end{eqnarray}%
where $L_{\eta _{j}}$ is the geometric derivative, that is, the Lie
derivative operator along $\eta _{j}$.

\subsection{A Special Case of $L_{1}~$}

\label{spec} Suppose that $L_{1}\left( t,x^{k},\dot{x}^{k}\right)
=L_{1}\left( t,x^{k}\right) =-V_{1}\left( t,x^{k}\right) .$ Then the Noether
symmetry conditions (\ref{ANS.07})-(\ref{ANS.09}) and (\ref{ANS.10})-(\ref%
{ANS.12}) transform to

$\left( \varepsilon \right) ^{0}:$%
\begin{eqnarray}
L_{\eta _{0}}g_{ij} &=&2\left( \frac{1}{2}\xi _{0,t}\right) g_{ij},
\label{ANS.13} \\
g_{ij}\eta _{0,t}^{i} &=&f_{0,j},  \label{ANS.14} \\
L_{\eta _{0}}V_{0}+\xi _{0,t}V_{0}+\xi _{0}V_{0\,,t}+f_{0,t} &=&0,
\label{ANS.15}
\end{eqnarray}

$\left( \varepsilon \right) ^{1}:$%
\begin{eqnarray}
L_{\eta _{1}}g_{ij} &=&2\left( \frac{1}{2}\xi _{1,t}\right) g_{ij},
\label{ANS.16} \\
g_{ij}\eta _{1,t}^{i} &=&f_{1,j},  \label{ANS.17} \\
L_{\eta _{0}}V_{1}+L_{\eta _{1}}V_{0}+\xi _{0}V_{1\,,t}+\xi _{0,t}V_{1}+\xi
_{1}V_{0\,,t}+\xi _{1,t}V_{0} &=&-f_{1,t},  \label{ANS.18}
\end{eqnarray}%
respectively.

In fact, with the use of equation (\ref{ANS.13}) and (\ref{ANS.16}), $\eta
_{A}^{i}$ can be expressed as $\eta _{A}^{i}=T_{A}\left( t\right)
Y_{A}^{i}\left( x^{k}\right) $ where the $Y_{A}^{i}$ are Homothetic vectors
(HVs) or Killing vectors (KVs), of the metric $g_{ij}$ with conformal factor 
$\psi $, and therefore $T_{A}L_{Y_{A}}g_{ij}=2T_{A}\psi _{A}g_{ij}=2\left( 
\frac{1}{2}\xi _{A,t}\right) g_{ij},~$so that we may specify the relation $%
\xi _{A,t}=2\psi _{A}T_{A}.$ Consequently, we have detected that the
approximate Noether symmetry conditions are

$\left( \varepsilon \right) ^{0}:$%
\begin{eqnarray}
L_{\eta _{0}}g_{ij} &=&2\psi _{0}g_{ij},  \label{ANS.19} \\
g_{ij}\eta _{0,t}^{i} &=&f_{0,j},  \label{ANS.20} \\
L_{\eta _{0}}V_{0}+2\psi _{0}T_{0}V_{0}+\xi _{0}V_{0\,,t}+f_{0,t} &=&0,
\label{ANS.21} \\
\xi _{0,t} &=&2\psi _{0}T_{0},  \label{ANS.22}
\end{eqnarray}

$\left( \varepsilon \right) ^{1}:$%
\begin{eqnarray}
L_{\eta _{1}}g_{ij} &=&2\psi _{1}g_{ij},  \label{ANS.23} \\
g_{ij}\eta _{1,t}^{i} &=&f_{1,j},  \label{ANS.24} \\
L_{\eta _{0}}V_{1}+L_{\eta _{1}}V_{0}+\xi _{0}V_{1\,,t}+2\psi
_{0}T_{0}V_{1}+\xi _{1}V_{0\,,t}+2\psi _{1}T_{1}V_{0} &=&-f_{1,t},
\label{ANS.25} \\
\xi _{1,t} &=&2\psi _{1}T_{1},  \label{ANS.26}
\end{eqnarray}%
which leads to the following theorem:

\begin{theorem}
\label{the1}The approximate Noether symmetries of the perturbed Lagrangian 
\begin{equation*}
L\left( t,x^{k},\dot{x}^{k},\varepsilon \right) =\frac{1}{2}g_{ij}\dot{x}^{i}%
\dot{x}^{j}-V_{0}\left( t,x^{k}\right) -\varepsilon V_{1}\left(
t,x^{k}\right) +O\left( \varepsilon ^{2}\right) ,
\end{equation*}%
are generated from the HV algebra of the metric $g_{ij}$.
\end{theorem}

This is an important result which tells us the maximum number of
approximate symmetries that can be constructed. Moreover, it implies that if
the metric tensor $g_{ij}$ does not admit any Homothetic/Killing vector
fields, then we will not be able to determine any approximate conservation laws
for the dynamical system.

\subsubsection{Approximate Noether Symmetries of $\left( \protect\varepsilon %
^{n}\right) $.}

For completeness we extend our analysis to the case of approximate
symmetries of order $\varepsilon ^{n}.$ Suppose we have a point
transformation of the form%
\begin{eqnarray}
\bar{x}^{i} &=&x^{i}+a\left( \eta _{\left( 0\right)
}^{i}+\dsum\limits_{\gamma =1}^{n}\varepsilon ^{\gamma }\eta _{\left( \gamma
\right) }^{i}+O\left( \varepsilon ^{\gamma +1}\right) \right) , \\
\bar{t} &=&t+a\left( \xi _{\left( 0\right) }+\dsum\limits_{\gamma
=1}^{n}\varepsilon ^{\gamma }\xi _{\left( \gamma \right) }^{i}+O\left(
\varepsilon ^{\gamma +1}\right) \right) ,
\end{eqnarray}%
with symmetry generator%
\begin{equation}
X=X_{0}+\dsum\limits_{\gamma =1}^{n}\varepsilon ^{\gamma }X_{\gamma
}+O\left( \varepsilon ^{\gamma +1}\right).
\end{equation}%

Hence, we find the generalized Noether condition 
\begin{eqnarray}
&&\left( X_{0}^{\left[ 1\right] }+\dsum\limits_{\gamma =1}^{n}\varepsilon
^{\gamma }X_{\gamma }^{\left[ 1\right] }\right) \left( L_{0}+\varepsilon
L_{1}\right) +\left( L_{0}+\varepsilon L_{1}\right) \frac{d}{dt}\left( \xi
_{0}+\dsum\limits_{\gamma =1}^{n}\varepsilon ^{\gamma }\xi _{\gamma }\right)
\notag \\
&&-\frac{d}{dt}\left( f_{0}+\dsum\limits_{\gamma =1}^{n}\varepsilon ^{\gamma
}f_{\gamma }\right) =O\left( \varepsilon ^{\gamma +1}\right) .
\end{eqnarray}%
The derivation of the Noether symmetry conditions is similar to that of the
previous section. For this reason, and for the sake of brevity we shall
not present further details, but merely state the relevant formulae. To this
end, the Noether symmetry conditions are

$\left( \varepsilon \right) ^{0}:$%
\begin{eqnarray}
L_{\eta _{0}}g_{ij} &=&2\left( \frac{1}{2}\xi _{0,t}\right) g_{ij},
\label{con1} \\
g_{ij}\eta _{0,t}^{i} &=&f_{0,j}, \\
L_{\eta _{0}}V_{0}+\xi _{0,t}V_{0}+\xi _{0}V_{0\,,t}+f_{0,t} &=&0,
\end{eqnarray}

$\left( \varepsilon \right) ^{\gamma }~,~\gamma =1\ldots n:$%
\begin{eqnarray}
L_{\eta _{\gamma -1}}h_{ij}+L_{\eta _{\gamma }}g_{ij}-2\left( \frac{1}{2}\xi
_{\gamma -1,t}\right) h_{ij}-2\left( \frac{1}{2}\xi _{\gamma ,t}\right)
g_{ij} &=&0, \\
h_{ij}\eta _{\gamma -1,t}^{i}+g_{ij}\eta _{\gamma ,t}^{i} &=&f_{\gamma ,j},
\\
L_{\eta _{\gamma -1}}V_{1}+L_{\eta \gamma }V_{0}+\xi _{\gamma
-1}V_{1\,,t}+\xi _{\gamma -1,t}V_{1}+\xi _{\gamma }V_{0\,,t}+\xi _{\gamma
,t}V_{0} &=&-f_{\gamma ,t}.  \label{con3}
\end{eqnarray}

An important observation is that Theorem \ref{the1} is still valid.

\section{Applications}

\label{section4}

The primary objective of our work was the development of explicit
approximate Noether symmetry conditions. Having accomplished this, we are
now in a position to apply the theory to some notorious examples from the
literature. Tracing the same progression so far, we start with some benign
1-dimensional cases, followed by 2-dimensional cases and finally, tackle the 
$n$-dimensional case. In every example, we have solved the Noether
conditions, but due to the economy of space, these calculations will not be
displayed here. Instead the Noether symmetries and some of the integrals are
simply listed below. The formulae for the derivation of the approximate
conservation laws are given in Appendix \ref{section3}.

\subsection{One-dimensional Perturbed Lagrangians}

%TCIMACRO{\TeXButton{B}{\begin{table}[tbp] \centering}}%
%BeginExpansion
\begin{table}[tbp] \centering%
%EndExpansion
\caption{Approximate symmetries for one-dimensional Perturbed Lagrangians}%
\begin{tabular}{|llc|}
\hline
\textrm{\textbf{Case}} & $L\left( t,x^{k},\dot{x}^{k},\varepsilon \right) $
& \textrm{\textbf{Noether Symmetry}} \\ \hline
(I) & $%
\begin{array}{lcl}
L_{0} & = & \frac{1}{2}\dot{x}^{2}+\omega _{0}\cos x, \\ 
L_{1} & = & \cos \left( \omega t\right) \cos x%
\end{array}%
$ & 1$^{st}$-order~$\left( X=X_{0}+\varepsilon X_{1}\right) $: $%
Z^{I}=\varepsilon \partial _{t}$ \\ 
&  & 2$^{nd}$-order $\left( X=X_{0}+\varepsilon X_{1}+\varepsilon
^{2}X_{2}\right) $: $Z^{I}=\varepsilon ^{2}\partial _{t}$ \\ \hline
(II) & $%
\begin{array}{lcl}
L_{0} & = & \frac{1}{2}\dot{x}^{2}+\frac{V_{0}}{x^{2}}, \\ 
L_{1} & = & \frac{V_{1}}{2t^{2}}x^{2}%
\end{array}%
$ & $%
\begin{array}{lcl}
Z_{1}^{II} & = & \varepsilon \left( 2t\partial _{t}+x\partial _{x}\right) 
\\ 
Z_{2}^{II} & = & \frac{1}{V_{1}}\partial _{t}+\varepsilon \left( 2\ln \left(
t\right) \partial _{t}+\frac{x}{t}\partial _{x}\right)  \\ 
Z_{3}^{II} & = & \varepsilon \left( t^{2}\partial _{t}+tx\partial
_{x}\right)  \\ 
Z_{4}^{II} & = & -\frac{1}{2V_{1}}t^{2}\partial _{t}-\frac{1}{2V_{1}}%
tx\partial _{x} \\ 
&  & +\varepsilon \left( t^{2}\left( \ln \left( t\right) -\frac{1}{2}\right)
\partial _{t}+t\ln \left( t\right) \partial _{x}\right)  \\ 
Z_{5}^{II} & = & 2t\partial _{t}+x\partial _{x} \\ 
Z_{6}^{II} & = & \varepsilon \partial _{t}%
\end{array}%
$ \\ \hline
(III) & $%
\begin{array}{lcl}
L_{0} & = & \frac{1}{2}\dot{x}^{2}-\frac{1}{2}x^{2}, \\ 
L_{1} & = & \frac{1}{2}e^{\omega t}x^{2}%
\end{array}%
$ & $%
\begin{array}{lcl}
Z_{1}^{III} & = & \left( \frac{4}{\omega }+\omega \right) \partial
_{t}+\varepsilon \left( \frac{2}{\omega }e^{\omega t}\partial
_{t}+c_{1}e^{\omega t}x\partial _{x}\right)  \\ 
Z_{2}^{III} & = & \varepsilon \left( \cos \left( 2t\right) \partial
_{t}+\sin \left( 2t\right) x\partial _{x}\right)  \\ 
Z_{3}^{III} & = & \varepsilon \left( \left( x\cos \left( 2t\right) +\sin
\left( 2t\right) \right) \partial _{x}\right)  \\ 
Z_{4}^{III} & = & \sin \left( 2t\right) \partial _{t}+2\omega \cos \left(
2t\right) x\partial _{x} \\ 
&  & +\varepsilon \left( e^{\omega t}\frac{2\left( \omega \sin \left(
2t\right) -2\cos \left( 2t\right) \right) }{\omega ^{2}+4}\partial
_{t}+e^{\omega t}\sin \left( 2t\right) \partial _{x}\right)  \\ 
Z_{5}^{III} & = & \cos \left( 2t\right) \partial _{t}-2\omega \sin \left(
2t\right) x\partial _{x} \\ 
&  & +\varepsilon \left( e^{^{\omega t}}\frac{2\left( \omega \cos \left(
2t\right) +2\sin \left( 2t\right) \right) }{\omega ^{2}+4}\partial
_{t}+e^{\omega t}\cos \left( 2t\right) \partial _{x}\right)  \\ 
Z_{6}^{III} & = & \varepsilon \left( \sin \left( t\right) \partial
_{x}\right)  \\ 
Z_{7}^{III} & = & \varepsilon \left( \cos \left( t\right) \partial
_{x}\right)  \\ 
Z_{8}^{III} & = & \omega \left( \omega \sin \left( t\right) +2\cos \left(
t\right) \right) \partial _{y}+\varepsilon \left( e^{\omega t}\sin \left(
t\right) \partial _{x}\right)  \\ 
Z_{9}^{III} & = & \omega \left( \omega \cos \left( t\right) -2\sin \left(
t\right) \right) \partial _{y}+\varepsilon \left( e^{\omega t}\cos
(t)\partial _{x}\right)  \\ 
Z_{10}^{III} & = & \varepsilon \left( \partial _{t}\right) 
\end{array}%
$ \\ \hline\hline
\end{tabular}
\label{tab001}%
%TCIMACRO{\TeXButton{E}{\end{table}}}%
%BeginExpansion
\end{table}%
%EndExpansion

Table \ref{tab001} summarizes the results of three different Lagrangians of
dimension one. The appropriate application of (\ref{ANS.07})--(\ref{ANS.12})
leads to the derivation of both exact and approximate Noether symmetries 
up-to first-order, whilst via conditions (\ref{con1}) - (\ref{con3}), we find
the approximate Noether symmetries up to second-order.

Case (I) corresponds to the Hamiltonian 
\begin{equation*}
H_{0}=\frac{1}{2}\dot{x}^{2}-\omega _{0}\cos x,\newline
~H_{1}=-\varepsilon \cos \left( \omega t\right) \cos x,
\end{equation*}%
while Case (II) possesses the Hamiltonian 
\begin{equation*}
H_{0}=\frac{1}{2}\dot{x}^{2}-\frac{V_{0}}{x^{2}}~,~\newline
H_{1}=-\varepsilon \frac{V_{1}}{2t^{2}}x^{2},
\end{equation*}%
and has the corresponding nonzero boundary terms: $f\left( Z_{2}^{II}\right)
=\varepsilon \frac{-x^{2}}{2t^{2}}$, $f\left( Z_{3}^{II}\right) =\varepsilon 
\frac{x^{2}}{2}$, $f\left( Z_{4}^{II}\right) =\frac{-x^{2}}{4V1}+\varepsilon
\left( \frac{x^{2}}{2}\left( \ln (t)+1\right) \right) $. 

Subsequently, the approximate Noether integrals are derived to be\footnote{%
An interesting discussion on the relation between perturbations and the $SL(2,R)~
$Lie algebra can be found in \cite{sl2r}.} 
\begin{eqnarray*}
I\left( Z_{1}^{II}\right)  &=&\varepsilon \left( 2tH_{0}-x\dot{x}\right) , \\
I\left( Z_{2}^{II}\right)  &=&\frac{1}{V_{1}}H_{0}+\varepsilon \left( \left(
2\ln \left( t\right) H_{0}-\frac{x}{t}\dot{x}-\frac{1}{2t^{2}}x^{2}\right) +%
\frac{1}{V_{1}}H_{1}\right) , \\
I\left( Z_{3}^{II}\right)  &=&\varepsilon \left( t^{2}H_{0}-tx\dot{x}+\frac{1%
}{2}x^{2}\right) , \\
I\left( Z_{4}^{II}\right)  &=&-\frac{1}{2V_{1}}t^{2}H_{0}+\frac{1}{2V_{1}}%
tx\partial _{x}-\frac{1}{4V_{1}}x^{2}+ \\
&&+\varepsilon \Bigg(\left( t^{2}\left( \ln \left( t\right) -\frac{1}{2}%
\right) H_{0}-t\ln \left( t\right) \dot{x}+\frac{1}{2}x^{2}\left( \ln \left(
t\right) +1\right) \right)  \\
&&-\frac{1}{2V_{1}}t^{2}H_{1}+\frac{1}{2V_{1}}tx\dot{x}\Bigg), \\
I\left( Z_{5}^{II}\right)  &=&2tH_{0}-x\dot{x}, \\
I\left( Z_{6}^{II}\right)  &=&\varepsilon H_{0}.
\end{eqnarray*}

On the other hand, Case (III) admits the Hamiltonian functions 
\begin{equation*}
H_{0}=\frac{1}{2}\dot{x}^{2}+\frac{1}{2}x^{2}~,~H_{1}=-\varepsilon \frac{1}{2%
}e^{\omega t}x^{2},
\end{equation*}%
while the corresponding boundary terms are%
\begin{eqnarray*}
f\left( Z_{1}^{III}\right) &=&\varepsilon \left( \frac{1}{2}\omega e^{\omega
t}x^{2}\right) , \\
f\left( Z_{2}^{III}\right) &=&\varepsilon \left( \frac{1}{2}\omega e^{\omega
t}x^{2}+x^{2}\cos \left( 2t\right) \right) , \\
f\left( Z_{3}^{III}\right) &=&\varepsilon \left( \frac{1}{2}\omega e^{\omega
t}x^{2}-x^{2}\sin \left( 2t\right) \right) , \\
f\left( Z_{4}^{III}\right) &=&-\omega x^{2}\sin \left( 2t\right)
+\varepsilon \left( \frac{1}{2}\omega e^{\omega t}x^{2}+\frac{1}{2}%
x^{2}e^{\omega t}\left( \omega \sin \left( 2t\right) +2\cos \left( 2t\right)
\right) \right) , \\
f\left( Z_{5}^{III}\right) &=&-\omega x^{2}\cos \left( 2t\right)
+\varepsilon \left( \frac{1}{2}\omega e^{\omega t}x^{2}+\frac{1}{2}%
x^{2}e^{\omega t}\left( \omega \cos \left( 2t\right) -2\sin \left( 2t\right)
\right) \right) , \\
f\left( Z_{6}^{III}\right) &=&\varepsilon \left( \frac{1}{2}\omega e^{\omega
t}x^{2}+x\cos \left( t\right) \right) , \\
f\left( Z_{7}^{III}\right) &=&\varepsilon \left( \frac{1}{2}\omega e^{\omega
t}x^{2}-x\sin \left( t\right) \right) , \\
f\left( Z_{8}^{III}\right) &=&x\omega \left( \omega \cos \left( t\right)
+2\sin \left( t\right) \right) +\varepsilon \left( \frac{1}{2}\omega
e^{\omega t}x^{2}+xe^{\omega t}\left( \omega \sin \left( t\right) +\cos
\left( t\right) \right) \right) , \\
f\left( Z_{9}^{III}\right) &=&-x\omega \left( \omega \sin \left( t\right)
+2\cos \left( t\right) \right) +\varepsilon \left( \frac{1}{2}\omega
e^{\omega t}x^{2}+xe^{\omega t}\left( \omega \cos \left( t\right) -\sin
\left( t\right) \right) \right) , \\
f\left( Z_{10}^{III}\right) &=&\varepsilon \left( \frac{1}{2}\omega
e^{\omega t}x^{2}\right) .
\end{eqnarray*}

\subsection{Two-Dimensional Perturbed Lagrangians}

%TCIMACRO{\TeXButton{B}{\begin{table}[tbp] \centering}}%
%BeginExpansion
\begin{table}[tbp] \centering%
%EndExpansion
\caption{Approximate symmetries for two-dimensional Perturbed Lagrangians}%
\begin{tabular}{|ccc|}
\hline
\textrm{\textbf{Case}} & $L\left( t,x^{k},\dot{x}^{k},\varepsilon \right) $
& \textrm{\textbf{Noether Symmetry}} \\ \hline
(IV) & $%
\begin{array}{lcl}
L_{0} & = & \frac{1}{2}\left( \dot{x}^{2}+\dot{y}^{2}\right)  \\ 
&  & -\frac{1}{2}\left( x^{2}+y^{2}\right)  \\ 
L_{1} & = & -x^{2}y+\frac{y^{3}}{3}%
\end{array}%
$ & $%
\begin{array}{lcl}
Z_{0}^{IV} & = & \partial _{t} \\ 
Z_{t}^{IV} & = & \varepsilon \partial _{t}~,~Z_{\frac{n\left( n-1\right) }{2}%
}=\varepsilon \left( y\partial x-x\partial _{y}\right) ~ \\ 
Z_{1,2}^{IV} & = & \varepsilon i\left( a_{2}e^{-i2t}-a_{1}e^{i2t}\right)
\partial _{t} \\ 
&  & +\left( a_{1}e^{i2t}+a_{2}e^{-i2t}\right) \left( x\partial
_{x}+y\partial _{x}\right)  \\ 
Z_{5,6}^{IV} & = & \varepsilon \left( a_{5}e^{it}+a_{6}e^{-it}\right) \left(
c_{1}\partial _{x}+c_{2}\partial _{y}\right) 
\end{array}%
$ \\ \hline
(V) & $%
\begin{array}{lcl}
L_{0} & = & \frac{1}{2}\left( \dot{x}^{2}+\dot{y}^{2}\right)  \\ 
&  & +\frac{1}{2}\frac{1}{x^{2}+y^{2}} \\ 
L_{1} & = & -\frac{1}{2}\left( x^{2}+y^{2}\right) 
\end{array}%
$ & $%
\begin{array}{lcl}
Z_{0}^{V} & = & \partial _{t} \\ 
Z_{t}^{V} & = & \varepsilon \partial _{t}~,~Z_{\frac{n\left( n-1\right) }{2}%
}=\varepsilon \left( y\partial x-x\partial _{y}\right) ~ \\ 
Z_{1}^{V} & = & \varepsilon \left( 2t\partial _{t}+x\partial _{x}+y\partial
_{y}\right)  \\ 
Z_{2}^{V} & = & -y\partial _{x}+x\partial _{y} \\ 
Z_{3}^{V} & = & \varepsilon \left( t^{2}\partial _{t}+tx\partial
_{x}+ty\partial _{y}\right)  \\ 
Z_{4}^{V} & = & -t/2~\partial _{t}-x/4~\partial _{x}-y/4~\partial _{y} \\ 
&  & +\varepsilon \left( t^{3}/3~\partial _{t}+xt^{2}/2~\partial
_{x}+yt^{2}/2~\partial _{y}\right)  \\ 
Z_{5}^{V} & = & -t^{2}/4~\partial _{t}-xt/4~\partial _{x}-yt/4~\partial _{y}
\\ 
&  & +\varepsilon \left( t^{4}/12~\partial _{t}+xt^{3}/6~\partial
_{x}+yt^{3}/6~\partial _{y}\right)  \\ 
&  & 
\end{array}%
$ \\ \hline\hline
\end{tabular}%
\label{tab0001}%
%TCIMACRO{\TeXButton{E}{\end{table}}}%
%BeginExpansion
\end{table}%
%EndExpansion

In Table \ref{tab0001}, we consider Case (IV) and (V), which are both two
dimensional approximate Lagrangians with the Hamiltonians 
\begin{equation*}
H_{0}=\frac{1}{2}\left( \dot{x}^{2}-\dot{y}^{2}\right) -\frac{1}{2}\left(
x^{2}+y^{2}\right) ~,~H_{1}=\varepsilon \left( x^{2}y-\frac{y^{3}}{3}\right)
,
\end{equation*}%
\begin{equation*}
H_{0}=\frac{1}{2}\left( \dot{x}^{2}+\dot{y}^{2}\right) +\frac{1}{2}\frac{1}{%
x^{2}+y^{2}}~,~H_{1}=\varepsilon \frac{1}{2}\left( x^{2}+y^{2}\right) ,
\end{equation*}%
respectively. 

For Case (IV) the Illustratively, the exact and approximate Noether
integrals of motion are%
\begin{eqnarray*}
I_{0} &=&H_{0}, \\
I_{t} &=&\varepsilon H_{0}~,~I_{\frac{n\left( n-1\right) }{2}}=\varepsilon
\left( y\dot{x}-x\dot{y}\right) , \\
I_{1,2} &=&\varepsilon \Bigg(i\left( a_{2}e^{-i2t}-a_{1}e^{i2t}\right)
E-\left( a_{1}e^{i2t}+a_{2}e^{-i2t}\right) \left( x\dot{x}+y\dot{y}\right) 
\\
&&+i\left( a_{1}e^{i2t}-a_{2}e^{-i2t}\right) \left( x^{2}+y^{2}\right) \Bigg)%
, \\
I_{5,6} &=&\varepsilon \left( \left( a_{5}e^{it}+a_{6}e^{-it}\right) \left(
c_{1}\dot{x}+c_{2}\dot{y}\right) -i\left( a_{5}e^{it}-a_{6}e^{-it}\right)
\left( c_{1}x+c_{2}y\right) \right) .
\end{eqnarray*}

We omit the derivation of the conservation laws for Case (V). They can be
calculated easily from the formulae (\ref{cone1}) and (\ref{cone2}).

\subsection{The $n-$Dimensional Perturbed Lagrangian}

As a final example, we culminate our results for the derivation of the $n$%
-dimensional flat case. Consider the approximate Lagrangian%
\begin{equation*}
L=\frac{1}{2}\delta _{ij}\dot{x}^{i}\dot{x}^{j}-\frac{1}{2}\left(
x^{i}x_{i}\right) -\varepsilon V_{1}\left( t,x^{j}\right) ,
\end{equation*}%
for a general function $V_{1}$.

The first-order $\left( X=X_{0}+\varepsilon X_{1}\right) $ approximate
Noether symmetries are 
\begin{eqnarray*}
Z_{t} &=&\varepsilon \partial _{t}~,~Z_{\frac{n\left( n-1\right) }{2}%
}=\varepsilon \mathbf{X}_{IJ},~ \\
Z_{1,2} &=&\varepsilon i\left( a_{2}e^{-i2t}-a_{1}e^{i2t}\right) \partial
_{t}+\left( a_{1}e^{i2t}+a_{2}e^{-i2t}\right) H^{,i}\partial _{i}, \\
Z_{5,6} &=&\varepsilon \left( a_{5}e^{it}+a_{6}e^{-it}\right) S_{J}^{,i}.
\end{eqnarray*}%
At this juncture, we are able to state that:

\begin{itemize}
\item $\mathbf{X}_{IJ}~$\ are the $\frac{n\left( n-1\right) }{2}$ rotations
of the Euclidean space,

\item $S_{J}^{i}$ are the $n$ gradient KVs of the Euclidean space,

\item $H^{i}$ is the HV of the Euclidean space.
\end{itemize}

Here, the Noether integrals are in fact only approximate in nature, namely%
\begin{eqnarray*}
I_{t} &=&\varepsilon H_{0}~,~~I_{\frac{n\left( n-1\right) }{2}}=\varepsilon 
\mathbf{X}_{IJ}\dot{x}^{i}\dot{x}^{j}, \\
I_{1,2} &=&\varepsilon \Bigg(i\left( a_{2}e^{-i2t}-a_{1}e^{i2t}\right)
H_{0}-\left( a_{1}e^{i2t}+a_{2}e^{-i2t}\right) H^{,i}\dot{x}_{i} \\
&&+i\left( a_{1}e^{i2t}-a_{2}e^{-i2t}\right) x^{i}x_{i}\Bigg), \\
I_{5,6} &=&\varepsilon \left( \left( a_{5}e^{it}+a_{6}e^{-it}\right)
S_{J}^{,i}\dot{x}_{i}-i\left( a_{5}e^{it}-a_{6}e^{-it}\right) x_{J}\right) ,
\end{eqnarray*}%
where $H_{0}$ is the Hamiltonian of the zero-order Lagrangian. It is
important to see that the exact integrals of the unperturbed system become
approximate integrals.

\section{Discussion}

\label{section5}

The main scope of this work was to devise  approximate Noether symmetry
conditions for regular perturbed Lagrangians in a geometric way,  to gain
understanding into the role of the underlying geometry in the existence of
approximate symmetries. The family of Lagrangians that we considered are
those that are defined by a Kinetic energy and a potential, which in general
describe the motion of a particle in a $n$-dimensional space under the
action of an autonomous force. 

We found a strong connection between the Homothetic algebra of the
underlying geometry and the approximated symmetries and in particular, in
the scenario where the perturbation terms do not modify the Kinetic energy, 
approximate symmetries  exist if and only if the metric that defines
the Kinetic energy admits a nontrivial Homothetic algebra. Finally, in order
to demonstrate our results, we determined in a geometric way, the approximate
symmetries for various systems of special interest. 

In a forthcoming work, we want to extend this analysis and perform various
classifications in which the perturbations are invariant under a specific
algebra.

\bigskip

\textbf{Acknowledgments:} AP acknowledges the financial support of FONDECYT
grant no. 3160121 and thanks the University of Athens for the hospitality
provided while this work was performed. SJ would like to acknowledge the
financial support from the National Research Foundation of South Africa
(99279). 

%TCIMACRO{\TeXButton{appendix}{\appendix}}%
%BeginExpansion
\appendix%
%EndExpansion

\section{Approximate conservation laws}

\label{section3}

In this appendix the approximate conservation law are determined as they are
given by Noether's theorem \cite{gov}. \ In general it is well known that
 if vector field $X=\xi \partial _{t}+\eta ^{i}\partial _{i},$~is a
symmetry for the Lagrangian $L=L\left( t,x^{k},\dot{x}^{k}\right) $ with a
boundary term $f$, then the following function is a conservation law:%
\begin{equation}
I=\xi H-\frac{\partial L}{\partial \dot{x}^{i}}\eta ^{i}+f,
\end{equation}%
where $H$ is the Hamiltonian function of $L.$

In a similar way, for any approximate Lagrangian $L=L_{0}+\varepsilon L_{1},~
$and an approximate Noether generator $X=X_{0}+\varepsilon X_{1},$ we derive
the exact conservation law, $I_{0}$ and the first order approximate
conservation law, $I_{1}$ as follows:%
\begin{eqnarray}
I_{0} &=&\xi _{0}H_{0}-\frac{\partial L_{0}}{\partial \dot{x}^{i}}\eta
_{0}^{i}+f_{0},  \label{cone1} \\
I_{1} &=&\left( H_{0}\xi _{1}-\frac{\partial L_{0}}{\partial \dot{x}^{i}}%
\eta _{1}^{i}+f_{1}\right) +\xi _{0}H_{1}-\frac{\partial L_{1}}{\partial 
\dot{x}^{i}}\eta _{0}^{i}.  \label{cone2}
\end{eqnarray}

A generalization of this idea to the higher-order case of $\varepsilon $,
with the symmetry generator 
\begin{equation}
X=X_{0}+\dsum\limits_{\gamma =1}^{n}\varepsilon ^{\gamma }X_{\gamma
}+O\left( \varepsilon ^{\gamma +1}\right) ,
\end{equation}%
leads us to deduce the formulae for the associated Noether integrals, viz. 
\begin{eqnarray}
I_{0} &=&\xi _{0}H_{0}-\frac{\partial L_{0}}{\partial \dot{x}^{i}}\eta
_{0}^{i}+f_{0}, \\
I_{\gamma } &=&\left( H_{0}\xi _{\gamma }-\frac{\partial L_{0}}{\partial 
\dot{x}^{i}}\eta _{\gamma }^{i}+f_{\gamma }\right) +\xi _{\gamma -1}H_{1}-%
\frac{\partial L_{1}}{\partial \dot{x}^{i}}\eta _{\gamma -1}^{i}.
\end{eqnarray}

%J. Comput. Appl. Math. 230 (1) (2009) 224Ð232

\end{document}